\documentclass[a4paper, 11pt, oneside]{article} 

\usepackage{geometry}
\usepackage{booktabs} 

\usepackage{mathptmx} 

\usepackage{fancyhdr}
\usepackage[normalem]{ulem}
\usepackage{microtype}
\usepackage{graphicx}
\usepackage{subfigure}
\usepackage{colortbl}
\definecolor{mygray}{gray}{.88}
\usepackage{multirow}
\usepackage{amsmath}
\usepackage{bm}
\usepackage{epsfig}
\usepackage{authblk}

\usepackage[hyphens]{url}
\usepackage{hyperref}
\usepackage{color}
\usepackage{soul}
\usepackage{bbding}
\definecolor{mygray}{gray}{.88}

\newcommand{\tabincell}[2]{\begin{tabular}{@{}#1@{}}#2\end{tabular}}

\begin{document} 

\begin{titlepage} 

	\centering 
	
	\scshape 
	
	\vspace*{\baselineskip} 
	
	
	\rule{\textwidth}{1.6pt}\vspace*{-\baselineskip}\vspace*{2pt} 
	\rule{\textwidth}{0.4pt} 
	
	\vspace{0.75\baselineskip} 
	
	{\LARGE AIBench: \\An Agile Domain-specific Benchmarking Methodology and an AI Benchmark Suite\\} 
	
	\vspace{0.75\baselineskip} 
	
	\rule{\textwidth}{0.4pt}\vspace*{-\baselineskip}\vspace{3.2pt} 
	\rule{\textwidth}{1.6pt} 
	
	\vspace{2\baselineskip} 
	
	
	
	\vspace*{3\baselineskip} 
	
	
	Abstract and Section 1 (Introduction) were contributed by Jianfeng Zhan. Section 2 was contributed by Jianfeng Zhan, Lei Wang, Wanling Gao, and Fei Tang. Section 3 was contributed by Jianfeng Zhan. Section 4.1 was contributed by Chunjie Luo, Fei Tang, Zihan Jiang, Wanling Gao, Jianfeng Zhan, and seventeen industry partners. Section 4.2 and component benchmarks were contributed by Wanling Gao, Chunjie Luo, Xingwang Xiong, Fei Tang, Zihan Jiang, Tianshu Hao, Fanda Fan, Xu Wen, Fan Zhang, Yunyou Huang, Jianan Chen, and Mengjia Du. Section 4.3 and micro benchmarks were contributed by Wanling Gao and Daoyi Zheng. Section 4.4 was contributed by Wanling Gao, Fei Tang, Lei Wang, and Jianfeng Zhan. Section 5 was contributed by Fei Tang, Wanling Gao, Lei Wang, and Jianfeng Zhan. Section 6 was contributed by Jianfeng Zhan, Wanling Gao, Fei Tang, Lei Wang, and Chuanxin Lan. Section 7 and Section 8 were contributed by Jianfeng Zhan, Wanling Gao, and Lei Wang. Rui Ren and Chen Zheng provide Testbed support.
	
	\vspace{0.5\baselineskip} 
	
	
	
	\vspace{0.5\baselineskip} 

	\vfill 
	
	
	\epsfig{file=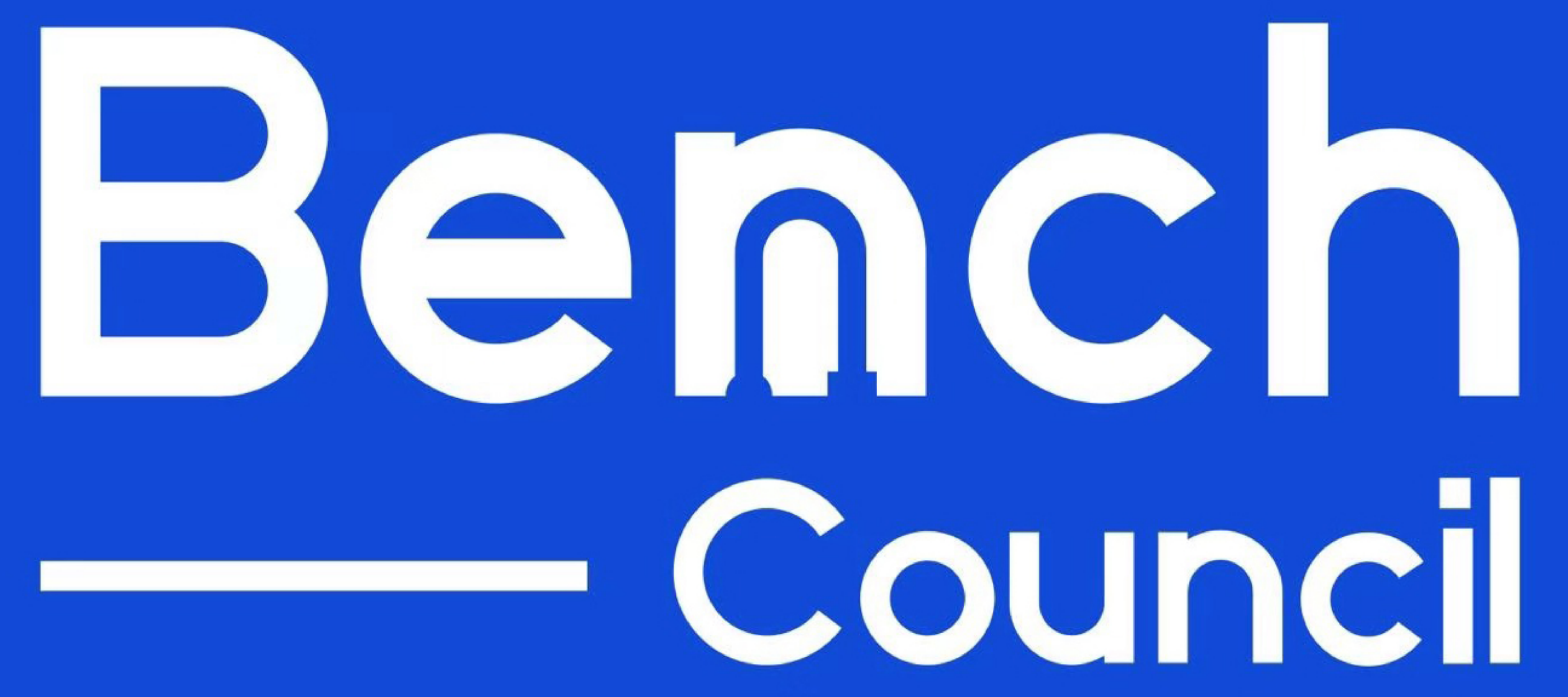,height=2cm}
	\textit{\\BenchCouncil: International Open Benchmarking Council\\Chinese Academy of Sciences\\Beijing, China\\http://www.benchcouncil.org/AIBench/index.html} 
	\vspace{5\baselineskip} 

	Technical Report No. BenchCouncil-AIBench-2020 
	
	{\large February 17, 2020} 

\end{titlepage}


\title{AIBench: An Agile Domain-specific Benchmarking Methodology and an AI Benchmark Suite}

\author[1,2,4]{Wanling Gao}
\author[1,4]{Fei Tang}
\author[1,2,4]{Jianfeng Zhan\thanks{Jianfeng Zhan is the corresponding author.}}
\author[1]{Chuanxin Lan}
\author[1,2,4]{Chunjie Luo}
\author[1,2,4]{Lei Wang}
\author[3]{Jiahui Dai}
\author[6]{Zheng Cao}
\author[1,4]{Xiongwang Xiong}
\author[1,4]{Zihan Jiang}
\author[1,4]{Tianshu Hao}
\author[1,4]{Fanda Fan}
\author[1,4]{Xu Wen}
\author[1,4]{Fan Zhang}
\author[1,4]{Yunyou Huang}
\author[1,4]{Jianan Chen}
\author[1,4]{Mengjia Du}
\author[1,4]{Rui Ren}
\author[1,2,4]{Chen Zheng}
\author[7]{Daoyi Zheng}
\author[8]{Haoning Tang}
\author[9]{Kunlin Zhan}
\author[10]{Biao Wang}
\author[11]{Defei Kong}
\author[12]{Minghe Yu}
\author[13]{Chongkang Tan}
\author[14]{Huan Li}
\author[15]{Xinhui Tian}
\author[16]{Yatao Li}
\author[17]{Gang Lu}
\author[18]{Junchao Shao}
\author[19]{Zhenyu Wang}
\author[20]{Xiaoyu Wang}
\author[3,5]{Hainan Ye}

\affil[1]{State Key Laboratory of Computer Architecture, Institute of Computing Technology, Chinese Academy of Sciences \\ \{gaowanling, tangfei, wanglei\_2011, zhanjianfeng, lanchuanxin\}@ict.ac.cn}
\affil[2]{BenchCouncil (International Open Benchmarking Council)}
\affil[3]{Beijing Academy of Frontier Sciences and Technology, \{daijiahui,yehainan\}@mail.bafst.com}
\affil[4]{University of Chinese Academy of Sciences}
\affil[5]{Xinxiu (SciCom)}
\affil[6]{Alibaba, zhengzhi.cz@alibaba-inc.com}
\affil[7]{Baidu, zhengdaoyi@baidu.com}
\affil[8]{Tencent, haoningtang@tencent.com}
\affil[9]{58.com, zhankunlin@58.com}
\affil[10]{NetEase, bjwangbiao@corp.netease.com}
\affil[11]{ByteDance, kongdefei@bytedance.com}
\affil[12]{Zhihu, yuminghe@zhihu.com}
\affil[13]{Lenovo, tanck1@lenovo.com}
\affil[14]{Paypal, huanli1@paypal.com}
\affil[15]{Moqi, xinhuit@moqi.ai}
\affil[16]{Microsoft Research Asia, yatli@microsoft.com}
\affil[17]{Huawei, lugang3@huawei.com}
\affil[18]{JD.com, shaojunchao@imdada.cn}
\affil[19]{CloudTa, wangzhenyu@cloudta.com.cn}
\affil[20]{Intellifusion, wang.xiaoyu@intellif.com}

\date{February 17, 2020}
\maketitle


\begin{abstract}

Domain-specific software and hardware co-design is encouraging as it is much easier to achieve efficiency for fewer tasks. 
Agile domain-specific benchmarking speeds up the process as it provides not only relevant design inputs but also relevant metrics, and tools. 
Unfortunately, modern workloads like Big data, AI, and Internet services dwarf the traditional one in terms of code size, deployment scale, and execution path, and hence raise serious benchmarking challenges.

This paper proposes an agile domain-specific benchmarking methodology. Together with seventeen  industry partners, we identify ten important end-to-end application scenarios, among which sixteen representative AI tasks are distilled as the AI component benchmarks.  We propose the permutations of essential AI and non-AI component benchmarks as end-to-end benchmarks. An end-to-end benchmark is a distillation of the essential attributes of an industry-scale application.  
We design and implement a highly extensible, configurable, and flexible benchmark framework, on the basis of which, we propose the guideline for building end-to-end benchmarks, and present the first end-to-end Internet service AI benchmark.

The preliminary evaluation shows the value of our benchmark suite---AIBench against MLPerf and TailBench for hardware and software designers, micro-architectural researchers, and code developers. 
The specifications, source code, testbed, and results are publicly available from the web site \url{http://www.benchcouncil.org/AIBench/index.html}.

\end{abstract}

\clearpage

\section{Introduction}

As it is much easier to achieve more efficient algorithms, systems, and architectures for fewer tasks, domain-specific software and hardware co-design is widely explored. For example, each of Internet service giants like Facebook, Google, Alibaba focuses on a specific application domain, i.e., search engine, social networks, E-commerce, respectively,  and they are active practitioners. The ongoing AI accelerator boom is another witness to this trend.   As the AI advancement has brought breakthroughs in processing images, video, speech, and audio ~\cite{lecun2015deep}, Internet service providers pervasively perform software and hardware AI co-design to augment their services~\cite{ni2018perceive,hazelwood2018applied,abadi2016tensorflow,jouppi2017datacenter,smith2017two}. This trend is also witnessed by  big data advancement, and there are hundreds of  single-purpose solutions in the forms of NoSQL, NewSQL or hardware accelerators. 

 Agile domain-specific benchmarking speeds up software and hardware co-design. Unfortunately, modern workloads dwarf the traditional one in terms of code size, deployment scale, and execution path, and hence raise serious benchmarking challenges. For example, the traditional desktop workloads, e.g., data compression~\cite{speccpu2017}, image manipulation~\cite{speccpu2017}, are about one hundred thousand lines of code, and run on a single node. The Web server workloads~\cite{nginx} are hundreds of thousands of lines of code, and run on a small scale cluster, i.e., dozens of nodes. However, for modern workloads, their runtime environment stacks (e.g., Spark~\cite{spark}, TensorFlow~\cite{abadi2016tensorflow}) alone are more than millions of lines of code, and these workloads often run on a large-scale cluster, i.e., tens of thousands of nodes~\cite{barroso2009datacenter}. Moreover, modern Internet services adopt a microservice-based architecture, which is often distributed across different datacenters, and consists of diversity of AI and non-AI modules with very long and complex execution paths. 
Worst of all, the real-world data sets, workloads or even AI models are hidden within the giant Internet service providers' datacenters~\cite{hazelwood2018applied,ayers2018memory}, which further exaggerates the benchmarking challenges. 
 
 On one hand,  the hardware and software designers should consider the overall system's effects.  Using  micro (interchangeable with kernel in this paper) or component benchmarks alone  can lead to incorrect conclusions. For example, in Section~\ref{eva_server}, we found that in terms of mere execution path,  end-to-end tail latency deteriorates even hundreds times comparing to a single AI component tail latency, which can not be predicted by a state-of-the-art statistical model~\cite{delimitrou2018amdahl} as discussed in Section~\ref{eva_server}. 
Hereby, end-to-end indicates the overall critical path. It may refer to the end-to-end (tail) latency of an online service,  or even cover offline AI training when updating an AI model for online services in a real time manner, as discussed in Section~\ref{modelupdate}.

 On the other hand,  it is usually difficult to justify
porting a full-scale end-to-end application
to a new computer system or architecture simply to obtain
a benchmark number~\cite{gao2018bigdatabench, bailey1991parallel}. For hardware designers, an end-to-end application is too huge to run on the simulators. In addition, evaluating a full-scale end-to-end application raises difficulties in reproducibility and interpretability of performance data~\cite{gao2018motif}, and may lead to an error-prone conclusion.     After gaining full knowledge of overall critical information, micro and component benchmarks  are still a necessary part of the evaluation.

 Put in other words, we believe a domain-specific benchmark suite should have three integrated parts. End-to-end benchmarks let software and hardware designer  learn about the overall system information. Each  end-to-end benchmark is a distillation of the essential attributes of an industry-scale application, and hence reduces the side effect of the latter's huge code size, extreme deployment scale, and complex execution paths.
 Measuring the achieved performance and quality targets for representative AI tasks, the component benchmarks   provides diverse computation and memory access patterns for the micro-architectural researchers. The   micro benchmarks are provided, and the code developers can drill down to hotspot functions for performance optimization.
 
 \begin{figure}[tb]
\centering
\includegraphics[scale=0.4]{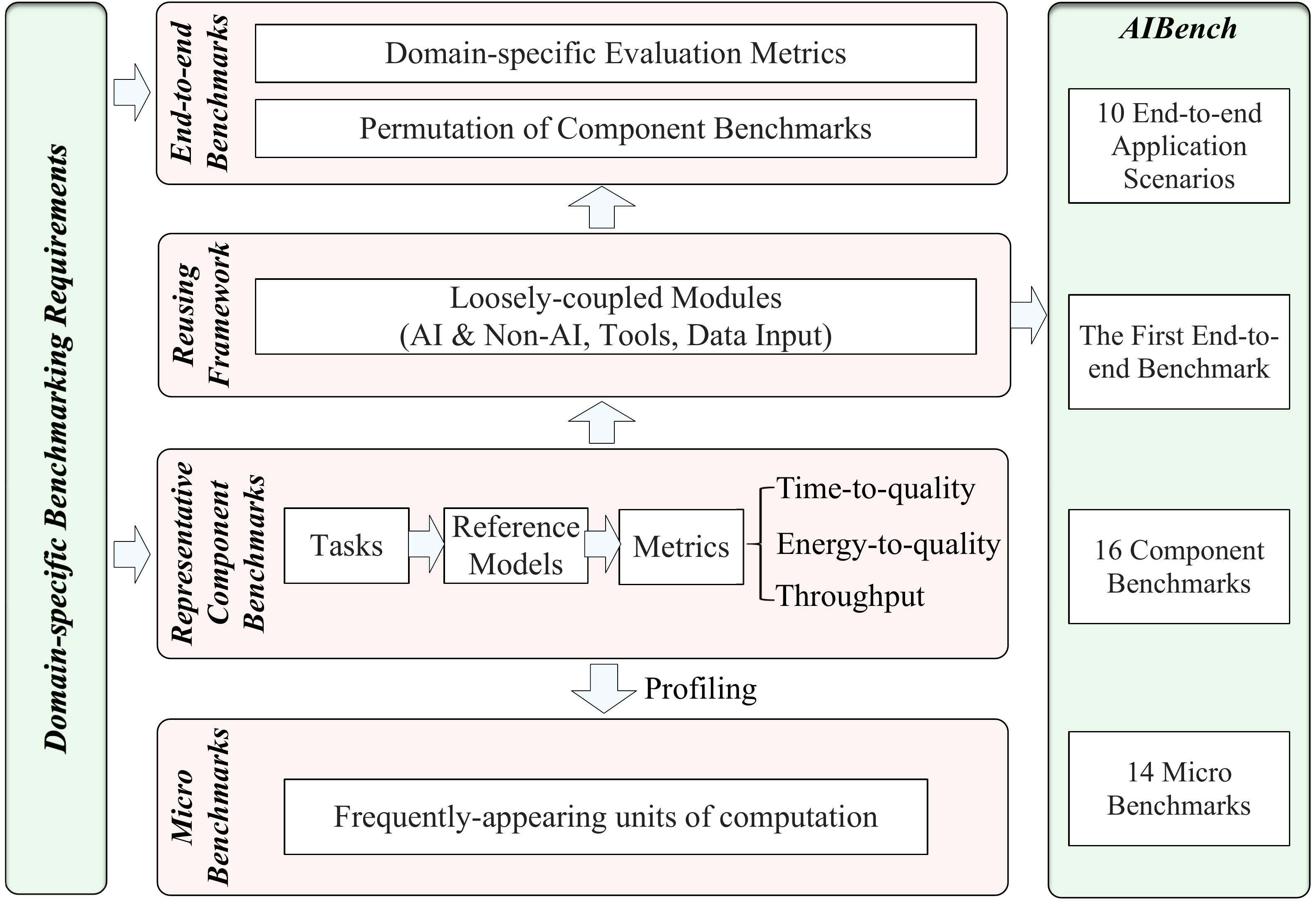}
\caption{The Agile Domain-specific Benchmarking Methodology.} 
\label{methodology1}
\end{figure}

This paper proposes an agile domain-specific benchmarking methodology as shown in Fig.~\ref{methodology1}. Without losing its generality, we apply it in characterizing the AI and Internet services application domains. First, in cooperation with seventeen  industry partners, we investigate their domain-specific benchmarking requirements, and extract ten important end-to-end application scenarios. Instead of using real-world applications, we propose the permutations of essential AI and non-AI tasks as end-to-end benchmarks.

Second, we identify  sixteen representative AI tasks as the AI component benchmarks with both performance and quality targets. After profiling sixteen AI component benchmarks, we identify and  implement  fourteen  frequent-appearing units of computation as the micro benchmarks.

Third, we present a highly extensible, configurable, and flexible benchmark framework,  
allowing  researchers to create end-to-end applications by using different components commonly found in major application domains.  On the basis of the framework, we propose guidelines on how to build end-to-end benchmarks, and design and implement the first end-to-end Internet service AI benchmark---E-commerce search intelligence.

The evaluation on a hybrid cluster consisting of 16-node CPUs and 4-node GPUs show the value of AIBench against MLPerf and TailBench. We gain many  insights for hardware and software designers, micro-architectural researchers, and code developers. Several important observations are as follows:  (1) In serving the same request, different AI components  incur significantly
different latency; an  end-to-end tail latency deteriorates
dozens times or even hundreds times with respect to a single
AI component, which can not be predicted by a state-of-the-art statistical model~\cite{delimitrou2018amdahl}. 
(2) Internet service architects must perform a tradeoff among  service quality, model complexity,  and model accuracy. (3)  AI models are updated in a real time manner in many end-to-end application scenarios. Offline training should be included into end-to-end benchmarking. (4) As they demonstrate distinct
computation and memory patterns, diverse AI tasks should be
included into the AI component benchmarks. (5) Drilling down to hotspot functions  is helpful for code optimization.

The rest of this paper is organized as follows. Section 2 explains the motivation.  Section 3 summarizes the  methodology. Section 4 describes how to characterize the AI and Internet service application domains. Section 5 illustrates how to build an end-to-end benchmark.
Section 6 performs evaluation. Section 7 summarizes the related work.
Section 8 draws a conclusion.

\section{Motivation}\label{moti}

\subsection{Why End-to-end Benchmarking Is Necessary}

Modern Internet services process millions of user queries daily, thus the tail latency is of paramount importance in terms of user experience~\cite{delimitrou2018amdahl}. However, a microservice-based architecture contains various AI and Non-AI modules， 
and consequently forms long and complex execution paths. Existing AI benchmarking efforts mostly provide a few micro or component benchmarks, and thus fail to model the critical paths and the permutation of primary components of an industry-scale application.

\textbf{The end-to-end tail latency deteriorates even 100X comparing to a single component tail latency.} The end-to-end tail latency indicates the overall performance of the entire execution path, while the component tail latency only reports the performance of a single module. Our experiments in Section~\ref{eva_server} show that the end-to-end tail latency deteriorates dozens times or even hundreds times comparing to a single component tail latency. For an AI component---recommendation, the difference is 13X, while for image classification, the difference reaches up to 296X.

Debugging the performance of a single component benchmark alone does not touch the full execution path and fail to provide bottleneck information among the primary modules within a critical path. Considering a 90th percentile latency, We found that among the four AI related components, the recommendation component occupies 72\% of the execution time, while the image classification component only  occupies 1.1\%. This  indicates that benchmarking a single AI component alone without the overall critical path does not make sense.

\subsection{Can a Statistical Model Predict the End-to-end Tail Latency?}

Someone may argue after profiling many components' tail latency performance, a statistical model can predict the end-to-end tail latency. Our answer is NO!
In Section~\ref{eva_server}, We use a state-of-the-art queuing theory~\cite{delimitrou2018amdahl} to evaluate the end-to-end application's latency and tail latency. Through the experimental evaluations, we find that the gap is 3.4 times between the actual average latency and the theoretical one, while the gap is 8.1 times between the actual 99th percentile latency and the theoretical one. Furthermore, the state-of-art queuing model~\cite{delimitrou2018amdahl} for tail latency takes the system as a whole, and is not suited for the end-to-end application that needs characterize the permutations of several or dozens of components.

\subsection{Why Offline AI Training is also Essential in End-to-end Benchmarking}

As witnessed by our many industry partners, when an AI model is used for online service, it has to be updated in a real time manner. For example, one E-commence giant  demands that the AI models have to be updated every one hour, and the updated model will bring in the award about 3\% click-through rate and millions of profits. In Section~\ref{modelupdate}, the evaluation shows offline training should be included into end-to-end benchmarking for performing tradeoffs among model update interval, training overhead, and accuracy improvement.

\section{Agile Domain-specific Benchmarking Methodology}\label{method}

As modern AI and Internet service workloads are not only diverse, but also fast changing and expanding, the traditional benchmark methodology that creates a new benchmark or proxy for every possible workload is prohibitively costly and even impossible~\cite{gao2018bigdatabench}. Hence an agile domain-specific benchmarking methodology is extremely essential.  Fig.~\ref{methodology1} summarizes our methodology. 

Step One. We investigate domain-specific benchmarking requirements with the industry partners. The input of this step is the candidate list of industry-scale applications. Just copying  the real-world applications is impossible for two reasons. First, they treat the real-world workloads, data sets, or models are confidential issues. Second, the massive code size, extreme deployment scale, and complex execution path make it infeasible.
So the purpose of this step is to understand their essential components and the permutation of different components.

Step Two. On the basis of the output from Step One, This step distills representative AI and non-AI tasks.  Different from traditional task, each AI task like image classification has both performance and quality targets~\cite{mattson2019mlperf}. Generally, an AI component specification defines a task in a high level language~\cite{zhan2019view}, only algorithmically in a paper-and-pencil approach~\cite{bailey1991parallel}. We implement each task as a component benchmark. The benchmark also provides a reference AI model, evaluation metrics, and state-of-the-art quality target~\cite{mattson2019mlperf}.

Step Three. According to the output of Step Two, we profile the full component benchmarks and drill down to frequently-appearing and time-consuming units of computation. We implement those units of computation as micro benchmarks. Micro benchmarks are easily portable to new architecture and system, and are beneficial to fine-grained profiling and tuning.

Step Four. According to the outputs of Steps One and Two, we design and implement a reusing benchmark framework, including AI and non-AI component library, the  data input,  online inference, offline training, and deployment tool modules.

Step Five. On the basis of the benchmark framework, we build end-to-end benchmarks. Each end-to-end benchmark models the permutation of several or tens of essential AI or non-AI components, reflecting complex interactions among different modules and depicting overall system's performance. In addition, we propose  domain-specific evaluation metrics.

\section{The AIBench Design and Implementation}\label{framework}

We first give a summary of the seventeen Industry Partners' benchmarking requirements, and then identify the representative AI tasks (component benchmarks and  micro benchmarks). Finally, we propose the reusing benchmark framework.

\subsection{Seventeen Industry Partners' Benchmarking Requirements}

Collaborating with seventeen industry partners whose domains include search engine, e-commerce, social network, news feed, video and etc, we extract the essential end-to-end application scenarios from their products or services.

The real-world applications are complex, and we only distill the permutations of primary AI and non-AI tasks. Table~\ref{requirement} summarizes the list of end-to-end application scenarios. 

For example, the first scenario in Table~\ref{requirement}---E-commerce search intelligence is extracted from an E-commerce giant. A user will be classified into different groups to provide personalized services. The results are ranked according to the relations between the queries and the products. And the ranking is adjusted by learning from the history query and hitting logs. The recommended products are also returned with the search results to the users.  We extract this industry-scale application into several AI tasks like classification,  learning to rank, recommendation, and non-AI tasks like query parsing, database operation, and indexing. Section~\ref{end-to-end} will describe how to implement this benchmark on the basis of the reusing framework described in Section~\ref{framework}.

In general, end-to-end benchmarks concern overall system's effects, including  quality-ensured response latency, tail latency, and latency-bounded throughput. A quality-ensured performance example is that a quality (e.g., accuracy) deviation  from the target  is within 2\%. Different application scenarios have domain-specific evaluation metrics.
For example, several scenarios require that the AI model is updated in a real time manner.

\begin{table*}[htbp]
\caption{Domain-specific Benchmarking Requirements}
\renewcommand\arraystretch{1.05}
\footnotesize
\label{requirement}
\center 
\begin{tabular}{|p{0.8in}|p{1.3in}|p{1in}|p{0.83in}|p{0.73in}|p{0.67in}|}
\hline
 \textbf{End to End Application Scenario} & \textbf{Involved AI Task} & \textbf{Involved Non-AI Task} & \textbf{Data} & \textbf{Metrics} & \textbf{Model Update  Frequency} \\

\hline
E-commerce search intelligence & Classification; Learning to rank;   Recommendation & Query parsing, Database operation, Indexing & User Data, Product data, Query data &  Precision, Recall, Latency & High\\
\hline
Language and dialogue translation & Text-to-Text translation; Speech recognition & Query parsing & Text, Speech & Accuracy, Latency & Low\\
\hline
Content-based image retrieval & Object detection; Classification; Spatial transformer;  Image-to-Text &Query parsing, Indexing, Sort & Image & Precision, Recall, Latency & High\\
\hline
Web searching & Text summarization;  Learning to rank;   Recommendation &Query parsing, Indexing, Crawler, Sort, Hash & Product data, Query data &  Precision, Recall, Latency & High\\
\hline
Facial authentication and payment & Face embedding; 3D face recognition; & Encryption & Face image & Accuracy , Latency & Low\\
\hline
News feed & Recommendation & Database operation, Sort, Basic statistics, Filter & Text &   Precision, Recall & High\\
\hline 
Photo translation & Classification; Spatial transformer;  Text-to-Text translation& Query parsing & Image, Text &  Accuracy, BLEU, Latency & Low\\
\hline
 Live streaming  & Image generation; Image-to-Image & Video codec, Video capture &Image & Latency & Low\\
\hline
 Video services  & Image compression; Video prediction &  Video codec &Video & Accuracy, Latency & Low \\
\hline
Online gaming & 3D object reconstruction; Image generation; Image-to-Image & Rendering & Image &  Latency &  Low\\ 
\hline

\end{tabular}
\end{table*}

\subsection{Representative AI Tasks}\label{identify}

To cover a wide spectrum of AI Tasks, we thoroughly analyze the end-to-end application scenarios shown in Table~\ref{requirement}. In total, we identify sixteen representative AI tasks. For each AI task, we implement it on TensorFow~\cite{abadi2016tensorflow} and PyTorch~\cite{pytorch} as the AI component benchmarks. Table~\ref{AIBench_component} summarizes the sixteen component benchmarks in AIBench.

\textbf{Classification.} This task is to extract different thematic classes within the input data like an image or text file. It is a typical task in Internet services or other application domains, and is widely used in multiple scenarios, like category prediction and spam detection.

\textbf{Image Generation.} This task aims to provide an unsupervised learning problem to mimic the distribution of data and generate images. The typical scenario  includes image resolution enhancement, which is used to generate high-resolution image.

\textbf{Text-to-Text Translation.} This task needs to translate a text from one language to another, which is the most important field of computational linguistics. It can be used to translate a search query and translate dialogue.

\textbf{Image-to-Text.} This task is to generate the description of an image automatically. It can be used to generate image caption or recognize optical character.

\textbf{Image-to-Image.} This task is to convert an image from one representation to another one. It can be used to synthesize the images with different facial ages and simulate virtual makeup. 

\textbf{Speech Recognition.} This task is to recognize and translate a spoken language into text. This task is beneficial for voice search and voice dialogue translation.

\textbf{Face Embedding.} This task is to transform a facial image into a vector in an embedding space. The typical scenarios are facial similarity analysis and face recognition.

\textbf{3D Face Recognition.} This task is to recognize the 3D facial information from multiple images from different angles. This task mainly focuses on three-dimensional images, and is beneficial to the facial similarity and facial authentication scenario.

\textbf{Object Detection.} This task is to detect the objects within an image. The typical scenarios include vertical search and video object detection.

\textbf{Recommendation.} This task is to provide recommendations. This task is widely used for advertise recommendation, community recommendation, or product recommendation.

\textbf{Video Prediction.} This task is to predict the future video frames through predicting previous frames transformation. The typical scenarios are video compression and video encoding, for efficient video storage and transmission.

\textbf{Image Compression.} This task is to compress the images and reduce the redundancy~\cite{toderici2017full}. The task is important for Internet services in terms of reducing data storage overhead and improving data transmission efficiency.

\textbf{3D Object Reconstruction.} This task is to predict and reconstruct 3D objects~\cite{yan2016perspective}. The typical scenarios are maps search, light field rendering, virtual reality, and online gaming.

\textbf{Text Summarization.} This  task is to generate a text  summary, which is important for search results preview, headline generation, and keyword discovery.

\textbf{Spatial Transformer.} This task is to perform spatial transformations~\cite{jaderberg2015spatial}. A typical scenario is space invariance image retrieval, so that an image can be retrieved even if it is extremely stretched.

\textbf{Learning to Rank.} This task is to learn the attributes of a searched content and rank the scores for the results, which is the key for a search engine service.


\begin{table*}[htbp]
\scriptsize
\caption{Component Benchmarks in AIBench.}
\renewcommand\arraystretch{1.4}
\label{AIBench_component}
\center 
\begin{tabular}{|p{0.63in}|p{1.4in}|p{2in}|p{1.6in}|}
\hline
\textbf{No.} & \textbf{Component Benchmark} & \textbf{Algorithm} & \textbf{Data Set}\\
\hline
DC-AI-C1 & Image classification & ResNet50~\cite{he2016deep} &  ImageNet~\cite{deng2009imagenet}, Cifar~\cite{krizhevsky2014cifar} \\
\hline
DC-AI-C2 & Image generation & WassersteinGAN~\cite{arjovsky2017wasserstein} &  LSUN~\cite{yu2015lsun} \\
\hline
DC-AI-C3 & Text-to-Text translation & Transformer~\cite{vaswani2017attention}  &  WMT English-German~\cite{wmt} \\
\hline
DC-AI-C4 & Image-to-Text & Neural Image Caption Model~\cite{vinyals2017show} &  Microsoft COCO~\cite{lin2014microsoft} \\
\hline
DC-AI-C5 & Image-to-Image & CycleGAN~\cite{zhu2017unpaired} &  Cityscapes~\cite{cordts2016cityscapes} \\
\hline
DC-AI-C6 & Speech recognition & DeepSpeech2~\cite{amodei2016deep} &  Librispeech~\cite{panayotov2015librispeech} \\
\hline
DC-AI-C7 & Face embedding & Facenet~\cite{schroff2015facenet} &  LFW~\cite{huang2008labeled}, VGGFace2~\cite{cao2018vggface2} \\
\hline
DC-AI-C8 & 3D Face Recognition & 3D face models~\cite{vieriu2015facial} &  77,715 samples from 253 face IDs\\
\hline
DC-AI-C9 & Object detection & Faster R-CNN~\cite{ren2015faster} &  Microsoft COCO~\cite{lin2014microsoft} \\
\hline
DC-AI-C10 & Recommendation & Neural collaborative filtering~\cite{he2017neural} &  MovieLens~\cite{harper2016movielens} \\
\hline
DC-AI-C11 & Video prediction & Motion-Focused predictive models~\cite{finn2016unsupervised} &  Robot pushing data set~\cite{finn2016unsupervised} \\
\hline
DC-AI-C12 & Image compression & Recurrent neural network~\cite{toderici2017full} &  ImageNet~\cite{deng2009imagenet} \\
\hline
DC-AI-C13 & 3D object reconstruction & Convolutional encoder-decoder network~\cite{yan2016perspective} &  ShapeNet Data set~\cite{chang2015shapenet}\\
\hline
DC-AI-C14 & Text summarization & Sequence-to-sequence model~\cite{nallapati2016abstractive} &  Gigaword data set~\cite{rush2017neural} \\
\hline
DC-AI-C15 & Spatial transformer & Spatial transformer networks~\cite{jaderberg2015spatial} &  MNIST~\cite{lecun2010mnist} \\
\hline
DC-AI-C16 & Learning to rank & Ranking distillation~\cite{tang2018ranking} &  Gowalla~\cite{cho2011friendship} \\
\hline

\end{tabular}
\end{table*}

The AI tasks concern both performance and quality targets. The primary metrics include the samples processed per second, the wall clock time to train a model achieving a target quality (Time-to-quality)~\cite{coleman2017dawnbench}, the wall clock time to train the specified  epochs,  quality-ensured throughput, and the energy consumption to train a model achieving a target quality (Energy-to-quality)~\cite{coleman2017dawnbench}.

\subsection{The AIBench Micro Benchmarks}

After profiling the sixteen component benchmarks, we identify fourteen frequently-appearing units of computation. They are Covolution, Fully connected, Relu, Sigmoid, Tanh, MaxPooling, AvgPooling, CosineNorm, BatchNorm, Dropout, Element-wise multipy, Softmax, Data arrangement, and Memcpy.  We implement them as a set of micro benchmarks using TensorFlow~\cite{abadi2016tensorflow} and Pthreads.

\subsection{The AIBench Framework}

\begin{figure}[tb]
\centering
\includegraphics[scale=0.8]{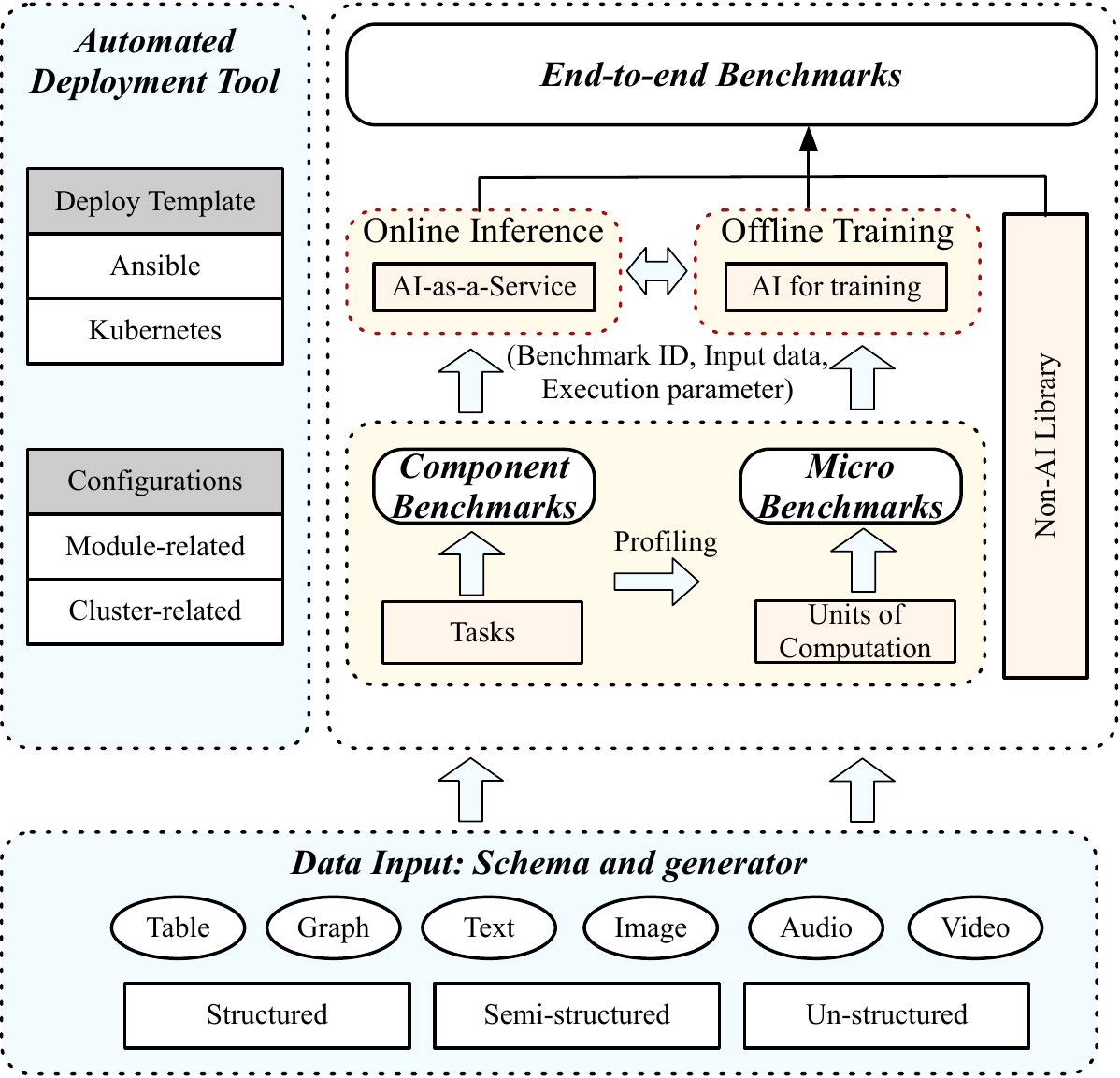}
\caption{Reusing Framework.} 
\label{arch-design}
\end{figure}

As shown in Fig.~\ref{arch-design}, the framework provides loosely coupled modules that can be easily configured. Currently, the AIBench framework includes  data input,  offline training, online inference, non-AI library,  and deployment tool modules. On the basis of the AIBench framework, we can easily  compose an end-to-end benchmark.

The data input module is responsible for  feeding data into the other modules. It collects representative real-world data sets, which are from  not only  the authoritative public websites but also our industry partners after anonymization.  The data schema is designed to maintain the real-world data characteristics, so as to alleviate the confidential issue. Based on the data schema, a series of data generators are further provided to support an large-scale data generation, like user or product information. To  cover a wide spectrum of data characteristics, we take diverse data types, e.g., structured, semi-structured, un-structured, and different data sources, e.g., table, graph, text, image, audio, video, into account. Our framework integrates various open-source data storage systems, and supports large-scale data generation and deployment~\cite{ming2014bdgs}.

The offline training and online inference modules are provided to build an end-to-end benchmark. First, the offline training module chooses one or more component benchmarks, through specifying the required benchmark ID, input data, and execution parameters like batch size. Then the offline training module trains a model and provides the trained model to the online inference module. The online inference module loads the trained model onto the serving system, i.e., TensorFlow serving. The non-AI library module provides the non-AI computations and database access, including query parsing, database operations, indexing, sort, crawler, hash, encryption, basic statistics, filter, video codec, video capture, and rendering.
For a complex end-to-end application, the online inference, the non-AI library, and the offline training modules together constitute an overall critical path.

To be easily deployed on a large-scale cluster, the framework provides deployment tools that contain two automated deployment templates using Ansible and Kubernetes. The Ansible templates support scalable deployment on physical or virtual machines, while the kubernetes templates are used to deploy on a container cluster. A configuration file needs to be specified for installation and deployment, including module parameters like a  chosen benchmark ID, input data, and the cluster parameters like nodes, memory, and network information. Through the deployment tools, a user doesn't need to know how to install and run each individual module.

\section{Building end-to-to benchmarks}

In this section, we illustrate how to build end-to-end benchmarks, and later discuss the guideline.

\subsection{The Design and Implementation of an E-commerce Search Intelligence}\label{end-to-end}

On the basis of the reusing framework, we implement the first end-to-end AI application benchmark---an E-commerce search intelligence (in short, E-commerce).  This benchmark models the complete use-case of a realistic E-commerce search intelligence, covering both text searching and image searching scenarios.

\begin{figure*}[tb]
\centering
\includegraphics[scale=0.65]{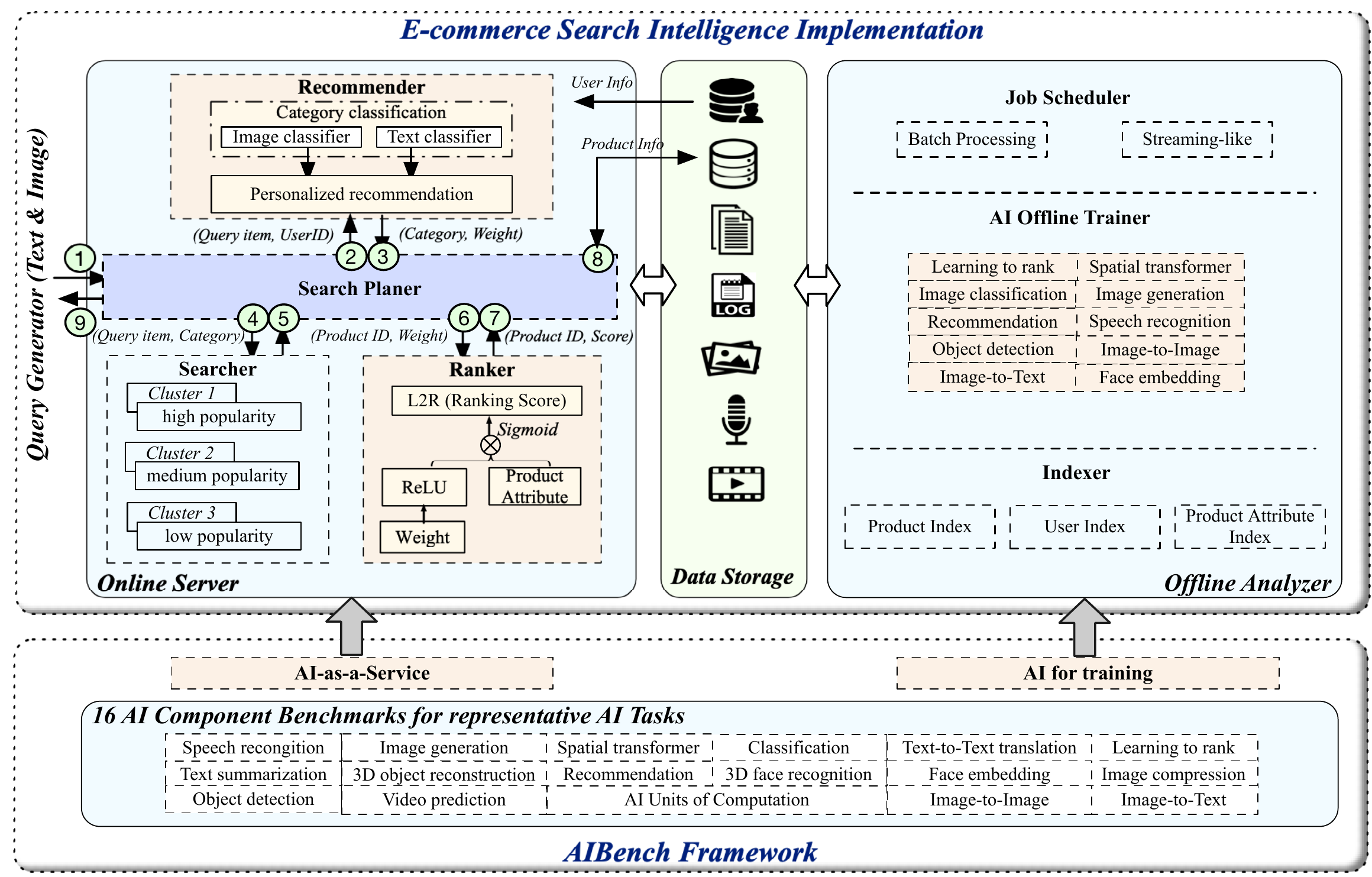}
\caption{AIBench Implementation.} 
\label{ecommerce-arch}
\end{figure*}

The E-commerce benchmark consists of four subsystems: online server, offline analyzer, query generator, and data storage, as shown in Fig.~\ref{ecommerce-arch}. Among them, online server receives the query requests and performs personalized searching and recommendation, integrating AI inference.

Offline analyzer chooses the appropriate AI component benchmarks and performs a training stage to generate a learning model. Also, offline analyzer is responsible to build data indexes to accelerate data access.

Query generator is to simulate concurrent users and send query requests to online server based on a specific configuration. Note that a query item provides either text or image to reflect different  search habits of users. The configuration designates the parameters like concurrency, query arriving rate,  distribution, user thinking time, and the ratio of text items and image items. The configurations simulate different query characteristics and satisfy multiple generation strategies. We implement our query generator based on JMeter~\cite{jmeter2017apache}.

The data storage module stores all kinds of data. The user database saves all the attributes of user information. The product database holds all the attributes of the product information. The logs record the complete query histories. The text data  contain the product description text or the user comments. The image and video data  depict the appearance and usage of product vividly. The audio data store the voice search data and voice chat data. Overall, the data storage covers various data types including structured, unstructured, and semi-structured data, and diverse data sources, including table, text, image, audio and video.

To support scalable deployment on the clusters with different scales, each module is scalable and can be deployed  on multiple nodes. Also, a series of data generators are provided to generate E-commerce data with different scales, through setting several parameters, e.g., the number of products and product attribute fields, the number of users and user attribute fields.

\subsubsection{Online Server}

Online server provides personalized searching and recommendations. 
Online server consists of four modules, including search planer, recommender, searcher, and ranker.

\textbf{Search planer} is the entrance of online server.  It is responsible for receiving the query requests from query generator, and sending the request to the other modules and receiving the return results. We use the Spring Boot framework~\cite{webb2013spring} to implement search planer.

\textbf{Recommender} is to analyze the query item and provide personalized recommendation, according to the user information obtained from the user database. It first conducts query spelling correction and query rewriting, and then it predicts the belonged category of the query item based on two classification models---FastText~\cite{joulin2016fasttext} and  ResNet50~\cite{he2016deep}. FastText is for text classification when a query item is text data, and ResNet50~\cite{he2016deep} is for image classification when a query item is an image. Using a deep neural network proposed by Alibaba ~\cite{ni2018perceive},  query planer then conducts an inference process and uses the offline trained model to provide personalized recommendation. It returns two vectors: one is the probability vector of the predicted categories,  and the other is the  user preference score vector of product attributes, such as the user preference for brand, color and etc. 
We use TensorFlow serving~\cite{olston2017tensorflow} to provide text classification, image classification, and online recommendation services.

To guarantee scalability and service efficiency, searcher follows an industry-scale architecture. \textbf{Searcher} is deployed on several different clusters, and three  clusters are the default configuration. The clusters hold the inverted indexes of product information in memory to guarantee high concurrency and low latency. According to the click-through rate and purchase rate, the products belong to three categories according to the popularity---high, medium, and low, and the proportion of data volume is 15\%, 50\%, and 50\%, respectively. Note that the high popularity category is a subset of the medium popularity category.  The indexes of products with different popularity are stored into the different clusters.  
Given a searching request, the searcher searches these three clusters one by one until reaching a specific amount. Generally, the cluster that holds low popularity products is rarely searched in a realistic scenario.
So for each category, searcher adopts different deployment strategies. The cluster for high popularity contains more nodes and more backups to guarantee the searching efficiency. While the cluster for low popularity deploys the least number of nodes and backups. We use Elasticsearch~\cite{gormley2015elasticsearch} to set up and manage the Searcher deploying on the three clusters.

\textbf{Ranker} uses the weight returned by \emph{recommender} as an initial weight, and ranks the scores of products through a personalized L2R neural network~\cite{ni2018perceive}. Ranker uses TensorFlow serving~\cite{olston2017tensorflow} to implement product ranking.

\subsubsection{Offline Analyzer}

Offline analyzer is responsible for training models and building indexes to improve the online serving performance. It consists of three modules---AI offline trainer, job scheduler, and indexer.

AI offline trainer is to train models using the data stored in the database. Offline trainer digests the features of the product data, e.g., text, image, audio, video. To power the efficiency of online server, Offline trainer chooses ten AI algorithms (component benchmarks) from the AIBench framework. The ten component benchmarks include  classification  for category prediction, recommendation  for personalized recommendation, learning to ranking  for result scoring and ranking, image-to-text  for image caption, image-to-image and image generation for image resolution enhancement, face embedding  for face detection within an image, spatial transformer  for image rotating and resizing, object detection  for detecting video data, and speech recognition for audio data recognition. 
 
Job scheduler provides two kinds of training mechanisms: batch processing and streaming-like processing. In a realistic scenario, some models need to be updated frequently. For example, when users search an item and click one product showed in the first page, the application will immediately train a new model based on the product that the users just clicked, and make new recommendations shown in the second page. Our benchmark implementations consider this situation, and adopt a streaming-like approach to updating the models every several seconds. For batch processing, trainer will update the models every several hours.

Indexer is to build indexes for product information. Indexer provides three kinds of indexes: the inverted indexes with a few fields of products for searching, the forward indexes with a few fields for ranking, and the forward indexes with a majority of fields for summary generation.

\subsection{Guidelines}

We are implementing other end-to-end benchmarks listed in Table~\ref{requirement}.
There are some guidelines. 

(1) Determine the essential AI and non-AI component benchmarks.

(2) For each component benchmark, find the valid input data  and the data input module.

(3) Determine the valid permutation of AI and non-AI components. 

(4) Specify the module-related configurations, i.e., benchmark ID, input data, execution parameters, Non-AI library, and cluster-related configurations, i.e., node, memory, and network information.

(5) Specify the deployment strategy and write the scripts for the automated deployment tool. 

(6) Train the AI models of the selected AI component
benchmarks using the offline training module, and transfer the trained models to the online inference module. 

\section{Evaluation}

This section summarizes our evaluation using AIBench end-to-end, component and micro benchmarks. In Section~\ref{eva-end}, we explain why end-to-end benchmarking is necessary for both online server and offline trainer, and gain several insights, which can not be found using MLPerf~\cite{mlperfweb} and TailBench~\cite{kasture2016tailbench}. In Section~\ref{diverse_task}, we characterize diverse and distinct computation and memory patterns of sixteen AI tasks, emphasizing the necessity of including diverse AI tasks for benchmarking, which is also ignored by MLPerf~\cite{mlperfweb}. In Section~\ref{drillfunction},  we drill down to the hotspot functions, and analyze their execution stalls.

\subsection{Experiment Setup}

\subsubsection{Node Configurations}

We perform experiments on a 16-node CPU and 4-node GPU cluster. 
All the nodes are connected with a 1 Gb Ethernet network. Each CPU node is equipped with two Xeon E5645 processors and 32 GB memory. Each processor contains six physical out-of-order cores. Hyper-Threading is disabled. The OS version of each node is Linux CentOS 6.9 with the Linux kernel version 3.11.10. The software versions are JDK 1.8.0, Python 3.6.8, and GCC 5.4, respectively.
We perform offline training on four Nvidia Titan XP GPUs. Every Titan XP owns 3840 Nvidia Cuda cores and 12 GB memory.

\subsubsection{Performance Data Collection}

We use the network time protocol (NTP)~\cite{mills1985network} for synchronizing cluster-wide clock. 
We use a profiling tool---Perf~\cite{de2010new} to collect the CPU micro-architectural data through the hardware performance monitoring counters (PMCs).
For GPU profiling, we use the Nvidia profiling toolkit---nvprof~\cite{nvprof} to track the running performance of GPU. To profile accuracy-ensured performance, we first adjust the parameters, e.g., batch size, to achieve the state-of-the-art quality target of that model on a given dataset, and then sample 1,000 epochs using the same parameter settings. For the GAN based model, whose accuracy is hard to measure, we set their parameters according to the referenced paper and reproduce the results.
We run each benchmark three times and report the average numbers.

\subsection{The Necessity of End-to-end Benchmarking}\label{eva-end}

This subsection demonstrates why end-to-to benchmarking is necessary for both online services and offline trainer in Section~\ref{eva_server} and Section~\ref{modelupdate}, respectively.

\begin{figure}[tb]
\centering
\includegraphics[scale=0.5]{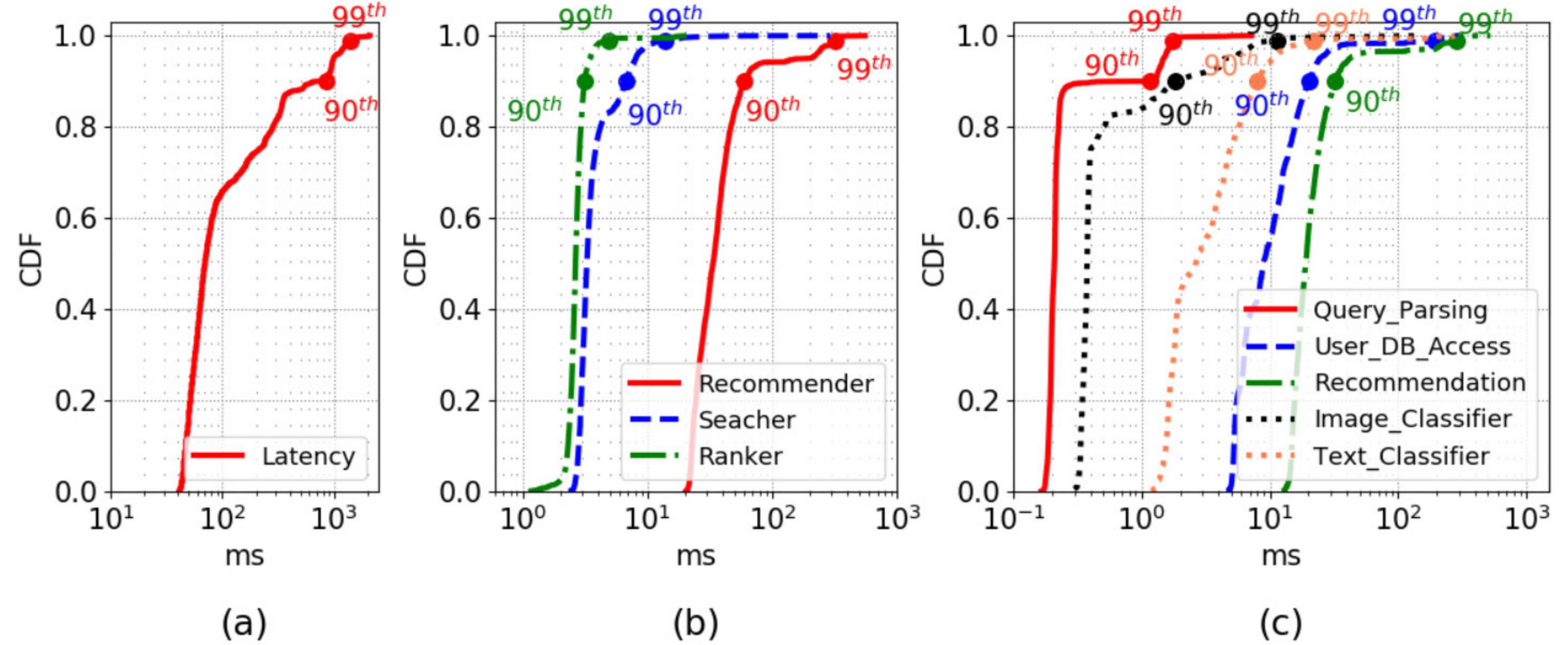}
\caption{Latency of Online Server.} 
\label{latency}
\end{figure}

\subsubsection{End-to-end Benchmarking is Necessary for Online Server}\label{eva_server}

We deploy online server on the 16-node CPU cluster.
Online server contains one query generator node (Jmeter 5.1.1), one search planer node (SpringBoot 2.1.3), two recommender nodes (TensorFlow Serving 1.14.0), nine searcher nodes (Elasticsearch 6.5.2), one ranker node (TensorFlow Serving 1.14.0), and two nodes for data storage (Neo4j 3.5.8 for the user database, Elasticsearch 6.5.2 for the product database).

The product database contains a hundred thousand products with 32-attribute fields. Query generator simulates 1000 users with 30-second warm up time. The users send query requests continuously every think time interval, which follow a  Poisson distribution. Note that the proportions of text queries and image queries are 90\% and 10\%, respectively. In total, we collect the performance numbers until 20,000 query requests have finished. We train each  AI task to achieve the quality target of the referenced paper. 

The latency is an important metric to evaluate the service quality. Fig.~\ref{latency}(a)~\footnote{With respect to the real numbers in our industry partner, the number is quite high. They have taken many measures to decrease the overall latency.} 
shows the end-to-end latency of online server. We find that the average, 90th percentile, and 99th percentile latency, of the entire execution path of the current  implementation is 215.5, 843, and 1419 milliseconds, respectively.

We further perform the latency breakdown of each module to identify the critical paths, including the recommender, searcher, search planer, and ranker modules, as shown in Fig.~\ref{latency}(b). The latency of  search planer is negligible, so we do not report it in Fig.~\ref{latency}(b). We find that recommender occupies the largest proportion of latency: 48 milliseconds, 60 milliseconds, and 317 milliseconds for the average, 90th percentile, 99th percentile latency, respectively. In comparison, the latency of searcher and ranker is both within 5 milliseconds, respectively. Although recommender and ranker both contain AI related components, they incur significantly different latencies.

Furthermore, Fig.~\ref{latency}(c) drills down the latency breakdown of the recommender module to a component level, which includes
query\_parsing, user\_DB\_access, image\_classifier, text\_classifier and recommendation. We find that user\_DB\_access (non-AI component) and recommendation (AI component) are the top two key components that impact the latency. Especially, the average latency of the recommendation component takes up 60\% of the average latency of the recommender module, and occupies 13\% of the total end-to-end latency of the online server subsystem. The 99th percentile latency of the recommendation component is 289 milliseconds, while the number for the recommender module and the whole subsystem are 317 milliseconds and 1419 milliseconds, respectively. The reason for that end-to-end tail latency deteriorates dozens times or even hundreds times with respect to a single component are 1) a single component may be not in the critical path; 2) even an AI component like recommendation is in the critical path, there exists cascading interaction effects with the other AI  and non-AI  components.

We also analyze the execution time ratio of the AI components vs. the non-AI  components in online server. If we exclude the data preprocessing and communication latency, the time spent on the AI components and the non-AI  components is 38 and 17 milliseconds for the average latency,  which indicates that the AI components are essential critical path of an industry-scale end-to-end benchmark like the E-commerce benchmark.

\textbf{Can a Statistical Model Predict the End-to-end Tail latency?}
As an end-to-end benchmark is much complex in using a hardware or software evaluation, an intuition is that can we use a statistical model to predict the end-to-end tail latency? The answer is NO!

The state-of-the-art work~\cite{delimitrou2018amdahl} uses the M/M/1 and M/M/K queuing models to calculate the p'th percentile latency. We repeat their work, and choose the M/M/1 model to predict the latency as we only deploy one  instance of online server. In the M/M/1 model, the p'th percentile latency ($T{p}$) and the average latency ($T{m}$) can be calculated using the following formula:
$T{p}=-\frac{\ln \left ( 1-\frac{P}{100} \right )}{\mu - \lambda }$
, $T{m}=\frac{1}{\mu - \lambda }$.
$\mu$ is the service rate, which follows the exponential distribution. $\lambda$ is the arrival rate, which follows the Poisson distribution.

We get the number of $\mu$---20 requests per second through the experiments. Then we set  $\lambda$ as 1.0 requests per second (10 simulated users), 9.1 requests per second (100 simulated users), and 16.7 requests per second (200 simulated users), respectively. For different settings, the theoretical number of the average latency is 53ms, 91ms, and 303ms, while the actual number is 123ms, 459ms, and 852ms, respectively. The average gap is 3.4 times. The theoretical number of the 99th percentile latency is 242ms, 422ms, and 1394ms, while  the actual number is 953ms, 5008ms, and 11980ms, respectively. The average gap is 8.1 times.  

The main reason for this huge gap is as follows. It is complex and uncertain to execute an end-to-end benchmark, and  the service rate  doesn't follow the exponential distribution. So, the M/M/1 model is far away from the realistic situation. However, the more generalized model (such as G/G/1 model) is difficult to be used to calculate the tail latency. Furthermore, if we try to characterize the permutations of executing dozens of components in an end-to-end benchmark, we need a more sophisticated analytical model such as a queuing network model, which is much infeasible to perform  a  calculation of tail latency.

\textbf{Tradeoff among Service Quality, Model Accuracy, and Model Complexity.}
The online inference module needs to load the trained model and conducts a forward computation to obtain the result. Usually, increasing the depth of a neural network model may improve the model accuracy, but it will lead to a larger model size and longer inference time. For comparison, we replace ResNet50 with ResNet152 in image\_classifier. The model accuracy improvement is 1.5\%, while the end-to-end 99th percentile latency deteriorates by 9.7X. 
Hence, Internet service architects must perform a tradeoff between the service quality, model complexity, and model accuracy.

\subsubsection{Tradeoff among Model Update Interval, Accuracy Improvement, and Training Overhead Using Offline Trainer}\label{modelupdate}

Updating AI models in a real time manner is a significant domain-specific concern in many scenarios. We evaluate the real-time model update efficiency using offline training. We deploy offline trainer on four Titan XP GPUs.

We adopt incremental learning method to update the models for online inference, and explore the relationship between the model update interval, training time overhead, and accuracy improvement. Our experiments show that comparing to the original training time and accuracy, 35\% additional training time brings in 1.9\% accuracy improvement for image\_classifier, and 10\% additional training time brings in 0.3\% accuracy improvement for ranker. 

Thus, offline training is an integrated part of end-to-end benchmarking. It not only facilitates measuring the model update efficiency, but also provides a guidance on how to choose an optimal update interval to balance the tradeoff between training overhead and accuracy improvement.

\subsection{Why Diversity of AI Tasks Matters for Benchmarking?}\label{diverse_task}

 We characterize distinct computation and memory patterns of the diverse AI tasks, emphasizing the necessity of including diverse AI tasks for benchmarking. 
 
 We characterize the sixteen component benchmarks of AIBench.
The AIBench component benchmarks are deployed on the Titan XP GPUs, and we focus on a single GPU performance. The CUDA and Nvidia driver versions are 10.0 and 410.78, respectively.

We evaluate the PyTorch implementations with the version of 1.1.0. The data set for each benchmark is as follows: ImageNet (137 GB) for image classification and Image compression; LSUN (42.8 GB) for image generation; VGGFace2 (36 GB) for face embedding; Microsoft COCO (13 GB) for Image-to-Text and object detection; MNIST (9.5 MB) for spatial transformer; Cityscapes (267 MB) for Image-to-Image; MovieLens (190 MB) for recommendation; Librispeech (59.3 GB) for speech recognition; Gowalla (107 MB) for learning to rank; WMT English-German (1.2 MB) for Text-to-Text translation; Robot pushing data set (137 GB) for Video prediction; ShapeNet Data set (6.8 GB) for 3D object reconstruction; Gigaword data set (277 MB) for Text summarization; 3D face data (37 GB) for 3D Face Recognition, respectively.

 GPU architecture contains multiple streaming multiprocessors (SM), each of which has a certain number of CUDA cores, memory registers, memory caches,  warp schedulers and etc. 
To characterize the AIBench component benchmarks from a perspectives of computation and memory access patterns, We choose five micro-architectural metrics, including achieved\_occupancy, ipc\_efficiency, gld\_efficiency, gst\_efficiency, and dram\_utilization. Achieved\_occupancy represents the ratio of the average active warps per active cycle to the maximum number of warps supported on a multiprocessor~\cite{nvprof}. 
Ipc\_efficiency indicates the ratio of the executed instructions per cycle to the theoretical number~\cite{nvprof}.
Gld\_efficiency means the ratio of the requested global memory load throughput to the required global memory load throughput~\cite{nvprof}.
Gst\_efficiency means the ratio of the requested global memory store throughput to the required global memory store throughput~\cite{nvprof}.
Dram\_utilization means the utilization level of the device memory relative to the peak utilization~\cite{nvprof}.

\begin{figure}[tb]
\centering
\includegraphics[scale=1]{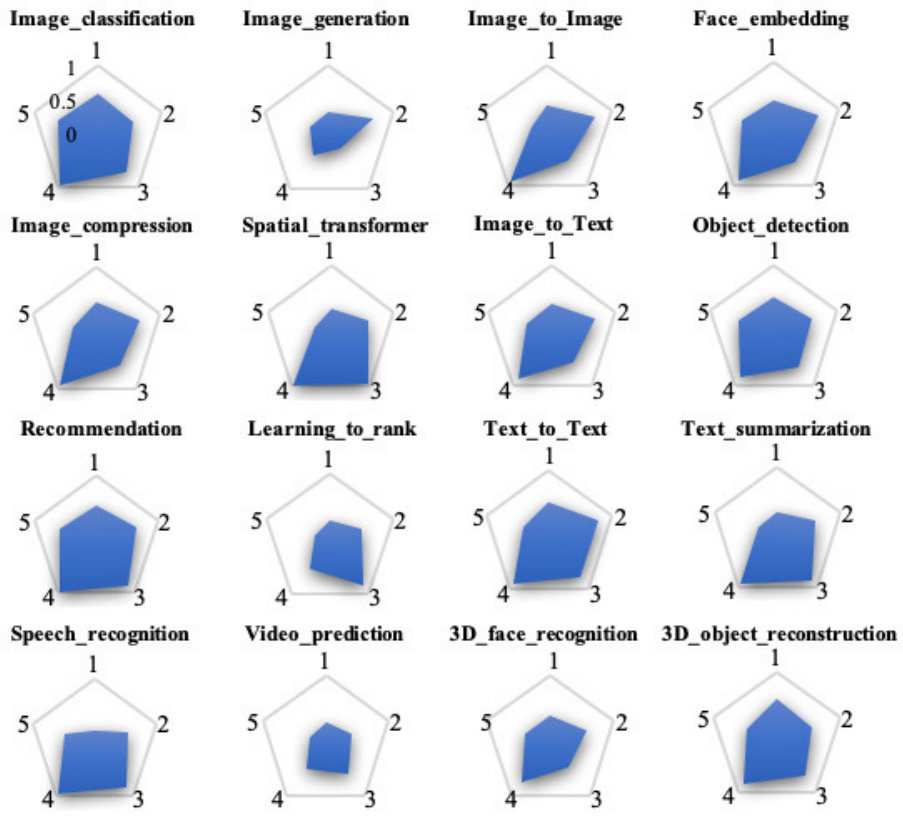}
\caption{Computation and Memory Patterns of AIBench Components (1: achieved\_occupancy; 2: ipc\_efficiency; 3: gld\_efficiency; 4: gst\_efficiency; 5: dram\_utilization). } 
\label{compumem}
\end{figure}

Fig. ~\ref{compumem} presents the computation and memory characteristics of the sixteen AI benchmarks. We find that they have distinct computation and memory patterns not only under different scenarios, e.g., processing text, image, audio, video, but also under different tasks of the same scenario, e.g., image classification and image generation. Thus,
diverse AI tasks reflecting different computation and memory access patterns should be included into the AI benchmarks. Achieving  a state-of-the-art quality target for each AI task will incur heavy training overhead, however, it does not justify including only a few benchmarks~\cite{zhan2019view}.

\subsection{Drill Down To Functional-level Code}~\label{drillfunction}

Following the experiments in~\ref{diverse_task}, We drill down to the hotspot functions and analyze their runtime breakdown and execution stalls for  code optimization.

The overall execution performance of these component benchmarks are varying in terms of IPC, which measures the executed instructions per cycle. From Fig.~\ref{compumem}, we find that the IPC efficiency ranges from 0.25 (Learning\_to\_rank) to 0.77  (Text\_to\_Text translation). Some benchmarks like learning\_to\_rank have extremely low IPC comparing to the other benchmarks. To discover the factors that impact the performance greatly, we first conduct runtime breakdown analysis and decompose the benchmarks into the hotspot kernels or functions, then we find the GPU execution efficiency in terms of different percentage of stalls.

\begin{figure}[tb]
\centering
\includegraphics[scale=0.8]{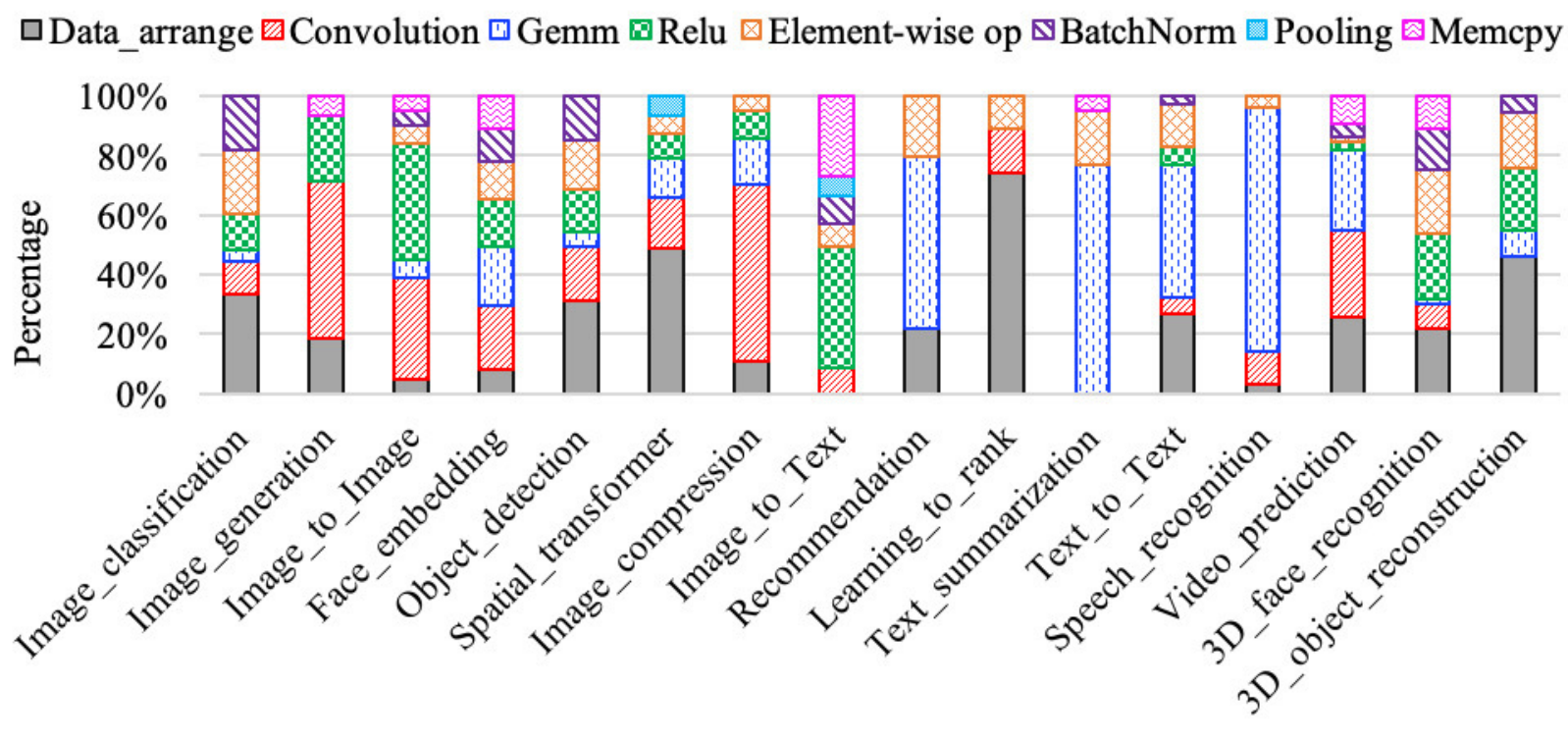}
\caption{Runtime Breakdown of AIBench Components.} 
\label{time-breakdown}
\end{figure}

\subsubsection{Runtime Breakdown}\label{timebreakdown} 

We use nvprof to trace the runtime breakdown and find the hotspot functions that occupy more than 80\% of runtime in total.
Since each run involves dozens of function calls, we single out the functions that occupy large proportions of runtime and classify them into several categories of kernels according to their computation logic.
Through statistics, we find that the most time-consuming functions among all  component benchmarks have much in common, and they are classified into eight categories of kernels, which are a subset of the AIBench micro benchmarks: data arrangement, convolution, general matrix multiply (gemm), batch normalization, element-wise operation, relu activation~\footnote{Relu activation is an element-wise operation, here we use a separate category of Relu considering its large proportion and diverse CUDA functions.}, pooling, and memory copy, spanning from computation kernels to memory access kernels.
Note that each kernel contains a bunch of functions that solve the similar issue. For example, a gemm kernel includes single or double precision floating general matrix multiply.
Fig.~\ref{time-breakdown} shows the runtime breakdown of sixteen component benchmarks, using the average number of all involved functions within each micro benchmark. Note that the remaining 20\% functions are not considered in this figure.
Further, for each micro benchmark, we summarize typical functions that occupy a large proportion of runtime among the component benchmarks, as shown in Table~\ref{func-sum}. We find that learning\_to\_rank spends too much time on data arrangement operations from Fig.~\ref{time-breakdown}, and the corresponding function call is maxwell\_scudnn\_128x32\_stridedB\_splitK\_interior\_nn with an IPC of 0.98. This is the reason why leaning\_to\_rank has the lowest IPC of 0.99.
We believe that the eight micro benchmarks and these corresponding functions are the optimization points not only for CUDA library optimizations but also for micro-architectural optimizations.

\begin{table}[htbp]
\scriptsize
\caption{Hotspot Functions.}
\renewcommand\arraystretch{1.2}
\scriptsize
\label{func-sum}
\center 
\begin{tabular}{|p{1.2in}|p{3in}|}
\hline
\textbf{Micro Benchmark} & \textbf{Function Name} \\
\hline
\multirow{3}{*}{Data Arragement} & maxwell\_scudnn\_128x128\_stridedB\_splitK\_interior\_nn \\
\cline{2-2}
& maxwell\_scudnn\_128x32\_stridedB\_splitK\_interior\_nn\\
\cline{2-2}
& maxwell\_scudnn\_128x128\_stridedB\_interior\_nn \\
\hline
\multirow{3}{*}{Convolution} &  maxwell\_scudnn\_winograd\_128x128\_ldg1\_ldg4\_tile148n\_nt \\
\cline{2-2}
& wgrad\_alg0\_engine \\
\cline{2-2}
 & fft2d\_r2c\_32x32 \\
\hline
\multirow{3}{*}{GEMM} & maxwell\_sgemm\_128x64\_nt \\
\cline{2-2}
& maxwell\_sgemm\_128x64\_nn \\
\cline{2-2}
& sgemm\_32x32x32\_NN\_vec \\
\hline
\multirow{3}{*}{BatchNorm} &  cudnn::detail::bn\_fw\_tr\_1C11\_kernel\_NCHW  \\
\cline{2-2}
& cudnn::detail::bn\_bw\_1C11\_kernel\_new  \\
\cline{2-2}
& batch\_norm\_backward\_kernel\\
\cline{2-2}
& at::native::batch\_norm\_backward\_kernel \\
\hline
\multirow{3}{*}{Relu} & maxwell\_scudnn\_128x128\_relu\_small\_nn  \\
\cline{2-2}
& maxwell\_scudnn\_128x128\_relu\_interior\_nn  \\
\cline{2-2}
& maxwell\_scudnn\_128x32\_relu\_interior\_nn \\
\hline
\multirow{3}{*}{\tabincell{l}{Element-wise}} & element-wise add kernel \\
\cline{2-2}
& element-wise threshold kernel \\
\cline{2-2}
& element-wise mul kernel \\
\hline
\multirow{2}{*}{Pooling} & MaxPoolBackward \\
\cline{2-2}
& AvePoolForward \\
\hline
\multirow{2}{*}{Memcpy} & CUDA memcpy HtoD \\
\cline{2-2}
& CUDA memcpy DtoD \\
\hline

\end{tabular}
\end{table}

\subsubsection{Stall Analysis} 

Focusing on the above eight most time-consuming micro benchmarks, we evaluate the following stalls of these kernels. Instruction fetch stall (Inst\_fetch) indicates the percentage of stalls because the next assembly instruction has not yet been fetched; Execution dependency stall (Exe\_depend) is the percentage of stalls because an input required by the instruction is not yet available; Memory dependency stall (Mem\_depend) is the percentage of stalls because a memory operation cannot be performed due to the required resources not being available or fully utilized; Texture stall (Texture) is the percentage of stalls because of the under-utilization of the texture sub-system;  Synchronization stall (Sync) is the percentage of stalls due to a syncthreads call; Constant memory dependency stall (Const\_mem\_depend) is the percentage of stalls because of immediate constant cache miss;  Pipe busy stall (Pipi\_busy) is percentage of stalls because a compute operation cannot be performed because the compute pipeline is busy; Memory throttle stall (Mem\_throttle) is the percentage of stalls due to large pending memory operations~\cite{nvprof}.

\begin{figure}[tb]
\centering
\includegraphics[scale=0.8]{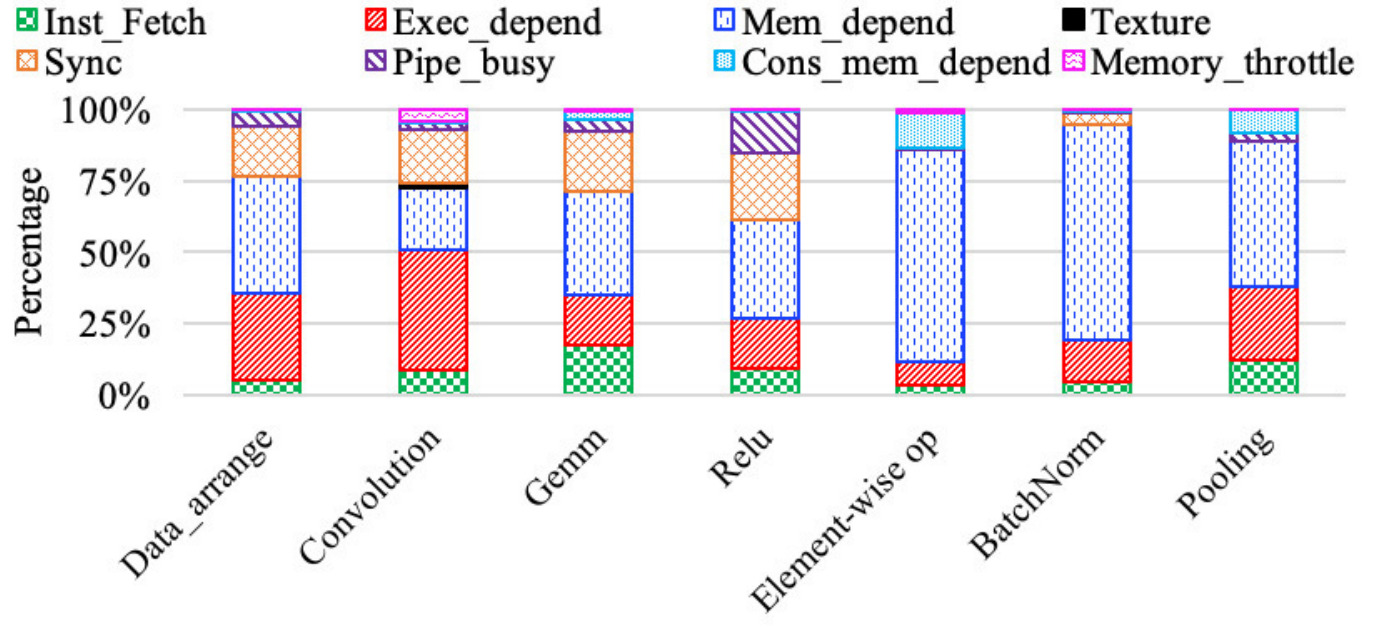}
\caption{Stall Breakdown of the Hotspot Functions.} 
\label{stall-breakdown}
\end{figure}

The breakdown of eight stalls of the hotspot functions is shown in Fig.~\ref{stall-breakdown}.
The top two GPU execution stalls are memory dependency stalls, and execution dependency stalls. For example, for Element-Wise benchmark, the memory dependency stalls occupy a very large proportion of 70\%, thus resulting in a low IPC number of about 0.86 on average.
The memory dependency stalls may occurs due to high cache misses, and thus the load/store resources are not available. Possible optimization strategies include optimizing date alignment, data locality, and data access patterns. The execution dependency stalls may occur due to low instruction-level parallelism, and exploiting ILP may alleviate partial execution dependency stalls to a certain degree.

\section{Related Work}

State-of-the-art and state-of-the-practise AI or Internet service benchmarks only provide a few micro or component benchmarks, as shown in Table~\ref{comparition_table}, and none of them distill representative and essential AI or non-AI components, and especially the permutations of different AI and non-AI components in characterizing industry-scale AI and Internet service applications.

\begin{table}[htbp]
\renewcommand\arraystretch{1.2}
\setlength{\abovecaptionskip}{0pt}
\scriptsize
\centering
\caption{AI Benchmark Comparison.}\label{comparition_table}
\center
\begin{tabular}{|p{0.6in}|p{0.3in}|p{0.5in}|p{0.5in}|p{0.5in}|p{0.5in}|p{0.5in}|p{0.6in}|p{0.4in}|}
\hline
\multicolumn{2}{|c|}{} & \tabincell{c}{AIBench} &\tabincell{c}{MLPerf} &\tabincell{c}{Fathom}
&\tabincell{c}{DeepBench}
&\tabincell{c}{DNNMark}
&\tabincell{c}{DAWNBench}
&\tabincell{c}{TBD}\\
\hline
\rowcolor{mygray} \multicolumn{9}{|l|}{Benchmark Framework (Extensible)}\\
\hline
\multicolumn{2}{|c|}{Modular-design} & \CheckmarkBold & $\times$ & $\times$ & $\times$ & \CheckmarkBold & $\times$ & $\times$\\
\hline
\rowcolor{mygray} \multicolumn{9}{|l|}{End-to-End Application Benchmark}\\
\hline
\multicolumn{2}{|c|}{Online module} & \CheckmarkBold & $\times$ & $\times$ & $\times$ & $\times$ & $\times$ & $\times$\\
\hline
\multicolumn{2}{|c|}{Offline module } & \CheckmarkBold & $\times$ & $\times$ & $\times$ & $\times$ & $\times$ & $\times$\\
\hline
\rowcolor{mygray} \multicolumn{9}{|l|}{Component Benchmark}\\
\hline
\multirow{2}{*}{\tabincell{l}{Image\\classification}} & \tabincell{l}{Train} & \CheckmarkBold & \CheckmarkBold & \CheckmarkBold & $\times$ & $\times$ & \CheckmarkBold & \CheckmarkBold \\
\cline{2-9}
& \tabincell{l}{Infer} & \CheckmarkBold & \CheckmarkBold & \CheckmarkBold & $\times$ & $\times$ & \CheckmarkBold &  $\times$ \\
\hline
\multirow{2}{*}{\tabincell{l}{Image\\generation}} & \tabincell{l}{Train} & \CheckmarkBold & $\times$ & $\times$ & $\times$ & $\times$ & $\times$ & \CheckmarkBold \\
\cline{2-9}
& \tabincell{l}{Infer} & \CheckmarkBold & $\times$ & $\times$ & $\times$ & $\times$ & $\times$ & $\times$\\
\hline
\multirow{2}{*}{\tabincell{l}{Text-to-Text}} & \tabincell{l}{Train} & \CheckmarkBold & \CheckmarkBold & \CheckmarkBold & $\times$ & $\times$ & $\times$ & \CheckmarkBold\\
\cline{2-9}
& \tabincell{l}{Infer} & \CheckmarkBold & \CheckmarkBold & \CheckmarkBold & $\times$ & $\times$ & $\times$ & $\times$\\
\hline
\multirow{2}{*}{\tabincell{l}{Image-to-Text}} & \tabincell{l}{Train} & \CheckmarkBold & $\times$ & $\times$ & $\times$ & $\times$ & $\times$ & $\times$\\
\cline{2-9}
& \tabincell{l}{Infer} & \CheckmarkBold & $\times$ & $\times$ & $\times$ & $\times$ & $\times$ & $\times$\\
\hline
\multirow{2}{*}{\tabincell{l}{Image-to-Image}} & \tabincell{l}{Train} & \CheckmarkBold & $\times$ & $\times$ & $\times$ & $\times$ & $\times$ & $\times$\\
\cline{2-9}
& \tabincell{l}{Infer} & \CheckmarkBold & $\times$ & $\times$ & $\times$ & $\times$ & $\times$ & $\times$\\
\hline
\multirow{2}{*}{\tabincell{l}{Speech recog-\\nition}} & \tabincell{l}{Train} & \CheckmarkBold & \CheckmarkBold & \CheckmarkBold & $\times$ & $\times$ & $\times$ & \CheckmarkBold\\
\cline{2-9}
& \tabincell{l}{Infer} & \CheckmarkBold & \CheckmarkBold & \CheckmarkBold & $\times$ & $\times$ & $\times$ & \CheckmarkBold\\
\hline
\multirow{2}{*}{\tabincell{l}{Face \\embedding}} & \tabincell{l}{Train} & \CheckmarkBold & $\times$ & $\times$ & $\times$ & $\times$ & $\times$ & $\times$\\
\cline{2-9}
& \tabincell{l}{Infer} & \CheckmarkBold & $\times$ & $\times$ & $\times$ & $\times$ & $\times$ & $\times$\\
\hline
\multirow{2}{*}{\tabincell{l}{3D Face \\Recognition}} & \tabincell{l}{Train} & \CheckmarkBold & $\times$ & $\times$ & $\times$ & $\times$ & $\times$ & $\times$\\
\cline{2-9}
& \tabincell{l}{Infer} & \CheckmarkBold & $\times$ & $\times$ & $\times$ & $\times$ & $\times$& $\times$\\
\hline
\multirow{2}{*}{\tabincell{l}{Object \\detection}} & \tabincell{l}{Train} & \CheckmarkBold & \CheckmarkBold & $\times$ & $\times$ & $\times$ & $\times$ & \CheckmarkBold \\
\cline{2-9}
& \tabincell{l}{Infer} & \CheckmarkBold & \CheckmarkBold & $\times$ & $\times$ & $\times$ & $\times$ & $\times$\\
\hline
\multirow{2}{*}{\tabincell{l}{Recommenda-\\tion}} & \tabincell{l}{Train} & \CheckmarkBold & \CheckmarkBold & $\times$ & $\times$ & $\times$ & $\times$ & \CheckmarkBold\\
\cline{2-9}
& \tabincell{l}{Infer} & \CheckmarkBold & $\times$ & $\times$ & $\times$ & $\times$ & $\times$& $\times$\\
\hline
\multirow{2}{*}{\tabincell{l}{Video \\prediction}} & \tabincell{l}{Train} & \CheckmarkBold & $\times$ & $\times$ &$\times$ & $\times$ & $\times$ & $\times$\\
\cline{2-9}
& \tabincell{l}{Infer} & \CheckmarkBold & $\times$ & $\times$ & $\times$ & $\times$ & $\times$ &$\times$\\
\hline
\multirow{2}{*}{\tabincell{l}{Image \\compression}} & \tabincell{l}{Train} & \CheckmarkBold & $\times$ & \CheckmarkBold & $\times$ & $\times$ & $\times$&$\times$ \\
\cline{2-9}
& \tabincell{l}{Infer} & \CheckmarkBold & $\times$ & \CheckmarkBold & $\times$ & $\times$ & $\times$ &$\times$\\
\hline
\multirow{2}{*}{\tabincell{l}{3D object re-\\construction}} & \tabincell{l}{Train} & \CheckmarkBold & $\times$ & $\times$ & $\times$ & $\times$ & $\times$ &$\times$\\
\cline{2-9}
& \tabincell{l}{Infer} & \CheckmarkBold & $\times$ & $\times$ & $\times$ & $\times$ & $\times$ &$\times$\\
\hline
\multirow{2}{*}{\tabincell{l}{Text sum-\\marization}} & \tabincell{l}{Train} & \CheckmarkBold & $\times$ & $\times$ & $\times$ & $\times$ & $\times$ &$\times$\\
\cline{2-9}
& \tabincell{l}{Infer} & \CheckmarkBold & $\times$ & $\times$ & $\times$ & $\times$ & $\times$ &$\times$\\
\hline
\multirow{2}{*}{\tabincell{l}{Spatial\\ transformer}} & \tabincell{l}{Train} & \CheckmarkBold & $\times$ & $\times$ & $\times$ & $\times$ & $\times$ &$\times$\\
\cline{2-9}
& \tabincell{l}{Infer} & \CheckmarkBold & $\times$ & $\times$ & $\times$ & $\times$ & $\times$ &$\times$\\
\hline
\multirow{2}{*}{\tabincell{l}{Learning to\\rank}} & \tabincell{l}{Train} & \CheckmarkBold & $\times$ & $\times$ & $\times$ & $\times$ & $\times$ &$\times$\\
\cline{2-9}
& \tabincell{l}{Infer} & \CheckmarkBold & $\times$ & $\times$ & $\times$ & $\times$ & $\times$& $\times$\\
\hline
\multirow{2}{*}{\tabincell{l}{Games}} & \tabincell{l}{Train} & $\times$ & \CheckmarkBold & \CheckmarkBold & $\times$ & $\times$ & $\times$ &\CheckmarkBold\\
\cline{2-9}
& \tabincell{l}{Infer} & $\times$ & $\times$ & \CheckmarkBold & $\times$ & $\times$ & $\times$ &$\times$\\
\hline
\multirow{2}{*}{\tabincell{l}{Memory \\network}} & \tabincell{l}{Train} & $\times$ & $\times$ & \CheckmarkBold & $\times$ & $\times$ & $\times$ & $\times$\\
\cline{2-9}
& \tabincell{l}{Infer} & $\times$ & $\times$ & \CheckmarkBold & $\times$ & $\times$ & $\times$ &$\times$\\
\hline
\multirow{2}{*}{\tabincell{l}{Question\\ answering}} & \tabincell{l}{Train} & $\times$ & $\times$ & $\times$ & $\times$ & $\times$ & \CheckmarkBold &$\times$\\
\cline{2-9}
& \tabincell{l}{Infer} & $\times$ & $\times$ & $\times$ & $\times$ & $\times$ & \CheckmarkBold &$\times$\\
\hline
\rowcolor{mygray} \multicolumn{9}{|l|}{Micro Benchmark}\\
\hline
\multicolumn{2}{|c|}{Convolution} & \CheckmarkBold & $\times$ & $\times$ & \CheckmarkBold & \CheckmarkBold & $\times$ & $\times$\\
\hline
\multicolumn{2}{|c|}{Fully connected} & \CheckmarkBold & $\times$ & $\times$ & \CheckmarkBold & \CheckmarkBold & $\times$ &$\times$\\
\hline
\multicolumn{2}{|c|}{\tabincell{c}{Element-wise op}} & \CheckmarkBold & $\times$ & $\times$ & $\times$ & $\times$ & $\times$ &$\times$\\
\hline
\multicolumn{2}{|c|}{Pooling} & \CheckmarkBold & $\times$ & $\times$ & $\times$ & \CheckmarkBold & $\times$ &$\times$\\
\hline
\multicolumn{2}{|c|}{Normalization} & \CheckmarkBold & $\times$ & $\times$ & $\times$ & \CheckmarkBold & $\times$ &$\times$\\
\hline
\multicolumn{2}{|c|}{Dropout} & \CheckmarkBold & $\times$ & $\times$ & $\times$ & \CheckmarkBold & $\times$ &$\times$\\
\hline
\multicolumn{2}{|c|}{Softmax} & \CheckmarkBold & $\times$ & $\times$ & $\times$ & \CheckmarkBold & $\times$ &$\times$\\
\hline
\multicolumn{2}{|c|}{Memory access} & \CheckmarkBold & $\times$ & $\times$ & $\times$ & $\times$ & $\times$ &$\times$\\
\hline
\multicolumn{2}{|c|}{AllReduce} & $\times$ & $\times$ & $\times$ & \CheckmarkBold & $\times$ & $\times$ &$\times$\\
\hline
\rowcolor{mygray} \multicolumn{9}{|l|}{Real-world Data sets and Software Stack}\\
\hline
\multicolumn{2}{|c|}{Text data} & 3 & 1 & 2 & N/A & N/A & 1 & 1\\
\hline
\multicolumn{2}{|c|}{Image data} & 8 & 2 & 2 & N/A & N/A & 2& 4 \\
\hline
\multicolumn{2}{|c|}{3D data} & 2 & 0 & 0 & N/A & N/A & 0 & 0\\
\hline
\multicolumn{2}{|c|}{Audio data} & 1 & 0 & 1 & N/A & N/A & 0 & 2\\
\hline
\multicolumn{2}{|c|}{Video data} & 1 & 0 & 1 & N/A & N/A & 0 & 0\\
\hline
\multicolumn{2}{|c|}{Software Stack} & 3 & 2 & 1 & 1 & 1 & 2 & 4\\
\hline

\end{tabular}
\end{table}

MLPerf~\cite{mlperf} is an ML benchmark suite targeting six AI tasks, including image classification, object detection, speech recognition, translation, recommendation, and reinforcement learning. It provides both light-weight and heavy-weight implementations.  Totally, it includes seven benchmarks for training and five benchmarks for inference.
The MLPerf training benchmark~\cite{mattson2019mlperf}  proposes  a series of benchmarking rules to eliminate the side effect of the stochastic nature of AI.

DAWNBench~\cite{coleman2017dawnbench} is a benchmark and competition focusing on end-to-end performance, which means the training or inference time to achieve a state-of-the-art accuracy. It only focuses on two component benchmarks including image classification on CIFAR10 and ImageNet, and question answering on SQuAD. 

Fathom~\cite{adolf2016fathom} provides eight deep learning component benchmarks implemented with TensorFlow.  Three of the eight benchmarks use different models for the image classification task. The Autoenc workload provides a variational autoencoder and can be used to reduce the dimension and compress images.

TBD Suite~\cite{zhu2018tbd} is a benchmark suite for DNN training. It provides eight neural network models that covers six AI tasks.  TailBench~\cite{kasture2016tailbench} is a benchmark suite consists of eight latency-critical workloads.

DeepBench~\cite{deepbench} consists of four operations involved in training deep neural networks, including three basic operations and recurrent layer operations. 
DNNMark~\cite{dong2017dnnmark} is a GPU benchmark suite that consists of a collection of deep neural network primitives.  Both DeepBench and DNNMark ignore the quality target in benchmarking.

Additionally, for machine learning and deep learning evaluation, MLModelScope~\cite{dakkak2019model} proposes a specification for repeatable model evaluation and a runtime to measure experiments.

There are two significant differences of AIBench from the other benchmark suite. One is to propose the permutations of essential AI and non-components as end-to-end benchmarks. We provide the reusing framework to speed up building end-to-end benchmarks.  The other is considering end-to-end benchmarks, components benchmarks and micro benchmarks as three integrated parts.  As a marked departure from the past, AIBench lets software and hardware designer  learn about the overall system information (end-to-end benchmarks),  provides diverse computation and memory access patterns (component benchmarks) as the design inputs for micro-architectural researchers, and drill down to hotspot functions (micro benchmarks) for the code
developers.

\section{Conclusion}

This paper proposes an agile domain-specific benchmarking methodology that speeds up software and hardware co-design.
Together with seventeen industry partners, we identify ten end-to-end application scenarios, distill sixteen representative AI tasks  and fourteen time-consuming units of computations. We propose the permutations of the essential AI and non-AI tasks as the end-to-end benchmark to characterize industry-scale applications. We design and implement a reusable framework to facilitate agile end-to-end benchmark building.  We build the first end-to-end benchmark to model E-commerce search intelligence. 
Our evaluation shows that the end-to-end benchmark integrating both online service and offline training provides overall system performance for hardware and software designers. The component benchmarks reflect diverse computation and memory access patterns, essential for  micro-architectural researchers. The micro benchmarks represent hotspot functions, beneficial to code optimization.






\end{document}